\documentclass[universe,article,pdftex,moreauthors]{Definitions/mdpi} 

\usepackage{xcolor}
\usepackage{natbib}
\firstpage{1} 
\pubvolume{1}
\issuenum{1}
\articlenumber{0}
\pubyear{2025}
\copyrightyear{2025}
\datereceived{ } 
\daterevised{ } 
\dateaccepted{ } 
\datepublished{ } 
\hreflink{https://doi.org/} 

\let\linenumbers\relax

\nolinenumbers

\Title{Re-examining Super-Nyquist Frequencies of 68 $\delta$ Scuti Stars Utilizing the $Kepler$ Long Cadence Photometry}

\TitleCitation{SNFs of 68 $\delta$ Scuti Stars}

\Author{Zilu Yang $^{1,2}$\orcidA{}, Jianning Fu $^{1,2,3}$*, Xuan Wang$^{1,2}$, Yanqi Mo$^{1,2}$ and Weikai Zong $^{1,2}$}

\AuthorNames{Zilu Yang, Jianning Fu, Xuan Wang, Yanqi Mo and Weikai Zong}

\AuthorCitation{Yang, Z.; Fu, J. ; Wang, X.; Mo, Y.; Zong, W.}

\address{%
$^{1}$ \quad Institute for Frontiers in Astronomy and Astrophysics, Beijing Normal University, Beijing 102206, China\\
$^{2}$ \quad School of Physics and Astronomy, Beijing Normal University, Beijing 100875, China\\
$^{3}$ \quad Xinjiang Astronomical Observatory, Chinese Academy of Sciences, Urumqi 830011, Xinjiang, China
}

\corres{Correspondence: jnfu@bnu.edu.cn}

\abstract{
The high-precision and long-duration photometry provided by the $Kepler$ mission has greatly advanced frequency analyses of a large number of pulsating stars, a fundamental step in asteroseismology.
For $\delta$ Scuti stars, analyses are typically confined to frequencies below the Nyquist frequency. 
However, signals above this limit can be reflected into the sub-Nyquist range, especially in long-cadence data, where they may overlap with genuine pulsation modes and lead to misinterpretation. 
To address this issue, a recently proposed method—the sliding Lomb–Scargle periodogram (sLSP)—can effectively distinguish real frequencies from aliased ones.
In this study, we compiled a sample of 68 $\delta$ Scuti stars whose frequency analyses were based on the $Kepler$ photometry. 
Using the sLSP method, we systematically examined the 1,406 reported frequencies in the literature. 
As a result, we identified 6 previously unrecognized reflected super-Nyquist frequencies in four stars: KIC 3440495, KIC 5709664, KIC 7368103, and KIC 9204718.
We have once again demonstrated the ability of the sLSP method to detect and correct such artifacts.
This technique improves the reliability of frequency selection, thereby enhancing the accuracy of asteroseismic interpretation and stellar modeling for the pulsating stars.
}


\begin{document}

\section{Introduction}

Asteroseismology investigates stellar evolution and internal structure by analyzing the pulsation frequencies of stars and constructing models based on these observations. 
Oscillations occur at various evolutionary stages and produce characteristic features in light curves, offering insights into internal processes such as rotation, diffusion, and convection \citep{2010aste.book.....A}.

$\delta$ Scuti ($\delta$ Sct) stars are a representative class of A- and F-type pulsating variables, positioned on or above the main sequence and concentrated toward the hotter end of the instability strip.
They are classified as p-mode pulsators, characterized by oscillations with frequencies $\gtrsim$ 4 $\mathrm{d}^{-1}$ (approximately 46 $\upmu$Hz), mainly driven by turbulent pressure and the $\kappa$-mechanism in the He\,\uppercase\expandafter{\romannumeral2} ionization zone of the stellar envelope \citep{2010aste.book.....A, 2022ARA&A..60...31K}.
High-precision, long-duration survey data show that most $\delta$ Sct stars also exhibit g-mode pulsations observed in $\gamma$ Doradus stars \citep{2011A&A...534A.125U, 2011MNRAS.415.3531B, 2014MNRAS.437.1476B, 2015MNRAS.452.3073B}.
Therefore, the complex pulsation behavior of $\delta$ Sct stars is influenced not only by the $\kappa$ mechanism but also by convective mechanisms that excite g-mode pulsations, such as convective blocking \citep{2005A&A...435..927D, 2015MNRAS.452.3073B, 2022ARA&A..60...31K}.
Modeling these oscillations offers an effective means to gain deeper insights into the $\delta$ Sct stars.

In recent years, space missions such as MOST \citep{2003PASP..115.1023W}, CoRoT \citep{2009A&A...506..411A}, the $Kepler$ \citep{2010Sci...327..977B}, and TESS \citep{2015JATIS...1a4003R} have achieved photometric precisions ranging from milli- to micro-magnitude levels.
The high-precision data significantly enhance the detection of low-amplitude oscillation frequencies, allowing for the identification of subtle signals that ground-based observations cannot capture, and enabling the construction of more complete frequency spectra.
This capability has directly contributed to recent advances in mode identification and the detection of complex oscillation patterns in various types of pulsating stars.
For instance, \citet{2022FrASS...932499L} proposed a method for diagnosing the non-linear pulsation characteristics of low-amplitude $\delta$ Sct stars, which may assist in identifying radial modes and determining their radial order $n$. 
\citet{2022ApJ...936...48Y} revealed the decline in the amplitude of the first-overtone mode in KIC 2857323, suggesting a possible loss of pulsation energy.
\citet{2024A&A...682L...8N} discovered significant variations in the amplitudes and frequencies of three independent pulsation modes, along with their harmonics and combinations, in KIC 6382916, challenging the reliability of independent modes as indicators of stellar interior structure.

To accurately map the internal sound-speed profile of pulsators through asteroseismology requires detecting a sufficient number of reliable oscillation frequencies.
The frequencies used in pulsation analysis are typically below the Nyquist frequency ($f_{\text{ny}}$), derived from Fourier transforms of photometric data.
Frequencies exceeding $f_{\text{ny}}$ are termed super-Nyquist frequencies (SNFs), and their reflected counterparts, $F_{\text{rsnf}}$, can alias with the real frequencies. 
Based on the $Kepler$ and TESS photometric observations, \citet{2013MNRAS.430.2986M} and \citet{2015MNRAS.453.2569M} outlined methods for identifying $F_{\text{rsnf}}$ through their spectral profiles. 
Besides aliasing, SNFs can cause false-alarm amplitude modulations in genuine pulsation modes, detectable by examining their characteristic modulation periods \citep{2016MNRAS.460.1970B, 2021RNAAS...5...41Z}. 
These spurious modulations resemble real pulsation features and must be properly interpreted to avoid mischaracterizing stellar behavior.

During the four-year $Kepler$ mission, SNFs are expected to undergo four full cycles of frequency modulation, providing a key signature to identify $F_{\text{rsnf}}$ \citep{2025A&A...693A..63W}.
Thus, \citet{2025A&A...693A..63W} introduced a sliding Lomb-Scargle periodogram (sLSP) method, which enables a more intuitive visualization of the sinusoidal-like modulation patterns of $F_{\text{rsnf}}$.
The $Kepler$ mission provides photometric data with two distinct sampling intervals: long-cadence (LC) data with a sampling interval of 29.43 minutes, and short-cadence (SC) data that sampled approximately every 58.85s, corresponding to $f_{\text{ny}}$ of 283.16 and 8496.18 $\upmu$Hz, respectively.
Limited by onboard storage and downlink constraints, $Kepler$ observed no more than 512 short cadence (SC) targets at a time, selected based on scientific priority \citep{2016ksci.rept....1V, 2010ApJ...713L.160G}.
Therefore, only a small fraction of the SC data is suitable for asteroseismic analysis.
However, SC data are essentially unaffected by SNFs, and thus can be used to verify the authenticity of SNFs detected in LC data \citep{2021AJ....161...27Y,2022RAA....22j5005F, 2024MNRAS.532.1140D}.
Additionally, in contrast to the limited SC observations, the longer continuous coverage of LC data effectively reduces the noise level \citep{2025A&A...693A..63W}. 
Moreover, the larger number of targets observed in LC mode offers significant support for frequency analysis.
In the study of \citet{2025A&A...693A..63W}, which analyzed $\gamma$ Doradus stars using the LC data, $F_{\text{rsnf}}$ was found to have only a minor impact on their g-mode pulsations, in contrast to its more significant influence on $\delta$ Sct stars.
This difference arises because the pulsation frequencies of $\delta$ Sct stars span the $f_{\text{ny}}$ of the $Kepler$ LC data.
SNFs map into the low-frequency domain, overlapping with the genuine pulsation frequencies \citep{2013MNRAS.430.2986M, 2025A&A...693A..63W}, which complicates the identification of intrinsic modes and can result in misclassifications during asteroseismic analysis \citep{2020MNRAS.498.1871M, 2022ApJ...936...48Y, 2024MNRAS.533.2705J, 2024AJ....168..171G}.

As the method proposed by \citet{2025A&A...693A..63W} can effectively remove $F_{\text{rsnf}}$ from the $Kepler$ data, we aim to re-examine the frequencies extracted from recent studies of $\delta$ Sct stars using the $Kepler$ data, with the goal of identifying any overlooked frequencies that may have slipped through the cracks.
Section 2 presents the selection of the research targets.
Section 3 describes the data preprocessing procedures and the subsequent frequency analysis. 
Section 4 provides a detailed discussion, and Section 5 summarizes the results and conclusion.

\section{Target Selection}

The $Kepler$ mission data, with its excellent frequency resolution and high duty cycle, have enabled the identification of over 2,000 $\delta$ Sct stars from its high-quality light curves \citep{2011MNRAS.417..591B,2014MNRAS.437.1476B,2016MNRAS.460.1970B, 2019A&A...630A.106G, 2025CoSka..55c.172L}.
We conducted a statistical review of research published over the past decades, which focus on the pulsation of the $\delta$ Sct stars based on Kepler observations, and identified a total of 74 publications in which pulsation frequencies were extracted and analyzed, covering 68 $\delta$ Sct stars. 
A detailed summary is provided in Table \ref{tab1}.

There are studies from the literature which have recognized the aliasing effects caused by SNFs and mitigated the impact of $F_{\text{rsnf}}$ by identifying multiplet structures or incorporating the SC data.
The 'Cad.' column in Table \ref{tab1} indicates the type of the $Kepler$ cadence used for frequency extraction in each study.
Since SNF primarily affects p-mode pulsations, we limit our analysis to significant frequencies above 46 $\upmu$Hz.
The 'Num' column lists the number of frequencies identified in each referenced study and used in our analysis.
If harmonics and combination frequencies were excluded in the study, the "Info" column is labeled as 'Inde.', indicating that only independent frequencies are counted in the 'Num' column. 
Otherwise, it is labeled as 'Sign.', meaning that all significant frequencies are included in the count.

\section{Data Preparation and Frequency Verification}

Given that the $f_{\text{ny}}$ of SC data (8496.18 $\upmu$Hz) is well above the pulsation range of $\delta$ Sct stars (typically confined below 800–900 $\upmu$Hz \citep{2010aste.book.....A, 2022ARA&A..60...31K}), and that even ground-based observations with very short exposure times \citep{1996A&A...307..529H, 2003MNRAS.341.1385K} have not detected frequencies beyond this range, we can be confident that aliasing with $F_{\text{rsnf}}$ does not occur in these cases.
Therefore, in the following work we focus primarily on the LC data.

\subsection{Preprocessing and Frequency Check}

We downloaded the $Kepler$ LC light curve data of the 68 $\delta$ Sct stars from the Mikulski Archive for Space Telescopes (MAST) server. 
To ensure high data quality, we utilized the Pre-search Data Conditioning Simple Aperture Photometry (PDCSAP) fluxes, which have been processed through the NASA $Kepler$ Data Processing Pipeline. 
This pipeline has effectively eliminated most discontinuities, outliers, systematic trends, and other instrumental signatures \citep{2012PASP..124.1000S}.
To further refine the data, we employed the Python package lightkurve \citep{2018ascl.soft12013L}, which allowed us to remove the discrete points from the PDCSAP flux, ensuring that only reliable data remained for analysis 
\citep{2024ApJS..271...57X}. 
Next, we applied the wotan package \citep{2019AJ....158..143H} to remove any remaining overall trends in the data. 
This method effectively removes long-term variations such as stellar activity or instrumental effects, thus ensuring that only the intrinsic pulsation signals are preserved for subsequent frequency analysis.
The operations were performed independently for each quarter of data, which were subsequently concatenated for further analysis.

Then, we applied the sLSP method to systematically examine 1,406 reported frequencies from 68 stars listed in Table \ref{tab1}. 
To achieve a balance between frequency resolution and smoothing effect, we adopted a sliding window of 300 days with a step size of 5 days, as proposed by \citet{2025A&A...693A..63W}.
Within each window, we computed the Lomb–Scargle periodogram (LSP) of the light curve to extract the corresponding frequency and amplitude information, thereby constructing the sLSP diagram.
Next, we selected data within a $\pm$0.1 $\upmu$Hz range around each target frequency to generate the LSP and sLSP plots, which reveal the amplitude distribution near the frequency and its temporal evolution, respectively.
Both the data processing and frequency extraction procedures may introduce slight deviations in the frequency values. 
If amplitude modulation is detected within $\pm$0.1 $\upmu$Hz of the target frequency, we consider it as a $F_{\text{rsnf}}$ candidate.

\begin{figure}[p]
\includegraphics[width=\textwidth]{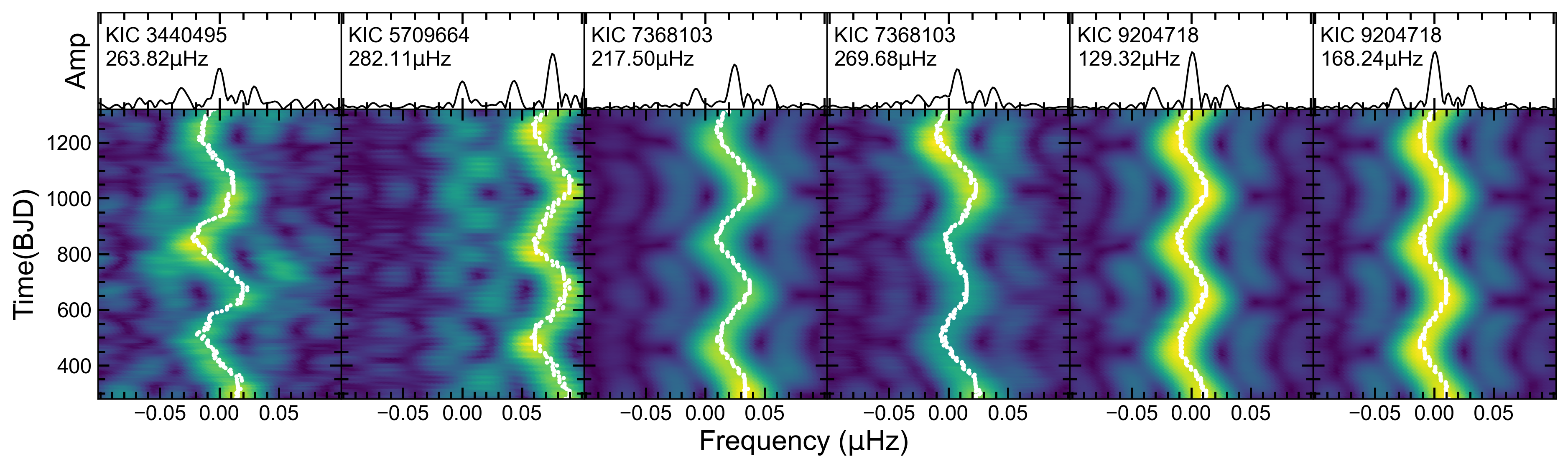}
\caption{The modulation of SNF that has not been discovered in the literature. The color intensity represents the amplitude of the corresponding frequency, with brighter colors indicating higher amplitudes.}\label{fig1}
\end{figure}

Ultimately, we found that the frequencies of four $\delta$ Sct stars exhibited SNF modulation, namely KIC 3440495 \citep{2022ApJ...937...80M}, KIC 5709664 \citep{2019MNRAS.486.2129D}, KIC 7368103 \citep{2019MNRAS.486.2462W} and KIC 9204718 \citep{2019NewA...71...33U}.
The SNFs modulation is illustrated in Figure \ref{fig1}.

\begin{table}[p]
\small
\caption{List of $\delta$ Sct stars whose pulsation frequency analysis was made with the $Kepler$ observations. 
The 'Cad.' column indicates the $Kepler$ cadence type used, with 'LSC' representing the use of both LC and SC data. 
The "Info" column is labeled 'Inde.' if harmonics and combination frequencies were excluded, or 'Sign.' if all significant frequencies were included.
The 'Num' column lists the corresponding number of frequencies identified in each study within the 46$\upmu$Hz–$f_{ny}$ range. '–' is used when no frequencies within this range were reported.
}\label{tab1}
\begin{tabular}{p{1.5cm} 
                >{\centering\arraybackslash}p{0.8cm} 
                >{\centering\arraybackslash}p{0.8cm} 
                >{\centering\arraybackslash}p{0.9cm} 
                >{\centering\arraybackslash}p{0.8cm}
                p{1.5cm} 
                >{\centering\arraybackslash}p{0.8cm} 
                >{\centering\arraybackslash}p{0.8cm} 
                >{\centering\arraybackslash}p{0.9cm} 
                >{\centering\arraybackslash}p{0.8cm}}
\toprule
\textbf{KIC ID} & \textbf{Cad.} & \textbf{Num} & \textbf{Info} & \textbf{Ref.} & \textbf{KIC ID} & \textbf{Cad.} & \textbf{Num} & \textbf{Info} & \textbf{Ref.} \\
\midrule

7914906 & LSC & 4 & Inde. & \mbox{\cite{2025PASJ...77..118L}}& 	5709664 & LC & 26 & Sign. & \mbox{\cite{2019MNRAS.486.2129D}} \\
2987660 & SC & 13 & Sign. & \mbox{\cite{2024NewA..11302294S}}& 	4142768 & LC & 78 & Inde. & \mbox{\cite{2019ApJ...885...46G}} \\
3429637 & LC & 16 & Inde. & \mbox{\cite{2024NewA..11302294S}}& 	4142768 & LC & 74 & Inde. & \mbox{\cite{2018MNRAS.476.4840B}} \\
3429637 & LSC & 7 & Sign. & \mbox{\cite{2012MNRAS.427.1418M}}& 	8113154 & LC & 28 & Inde. & \mbox{\cite{2019ApJ...884..165Z}} \\
4851217 & LC & 46 & Inde. & \mbox{\cite{2024MNRAS.533.2705J}}& 	10284901 & SC & 5 & Inde. & \mbox{\cite{2019ApJ...879...59Y}} \\
4851217 & SC & 4 & Inde. & \mbox{\cite{2020A&A...642A..91L}}& 	3441784 & SC & 30 & Inde. & \mbox{\cite{2019AJ....158...88A}} \\
9408694 & SC & 20 & Inde. & \mbox{\cite{2024MNRAS.532.1140D}}& 	5123889 & LC & 6 & Inde. & \mbox{\cite{2018MNRAS.475..359B}} \\
9408694 & LSC & 15 & Inde. & \mbox{\cite{2012MNRAS.419.3028B}}& 	6048106 & LC & 35 & Inde. & \mbox{\cite{2018AcA....68..425S}} \\
9851944 & SC & 33 & Inde. & \mbox{\cite{2024MNRAS.527.4052J}}& 	8553788 & SC & - & None & \mbox{\cite{2018A&A...616A.130L}} \\
9851944 & LC & 41 & Inde. & \mbox{\cite{2016ApJ...826...69G}}& 	8197761 & LC & 11 & Sign. & \mbox{\cite{2017MNRAS.467.4663S}} \\
10407873 & LC & 1 & Inde. & \mbox{\cite{2024ApJ...977..241W}}& 	9592855 & LC & 32 & Inde. & \mbox{\cite{2017ApJ...851...39G}} \\
10855535 & LC & 1 & Inde. & \mbox{\cite{2024ApJ...977...47S}}& 	10989032 & LC & - & None & \mbox{\cite{2017ApJ...850..125Z}} \\
8840638 & SC & 2 & Inde. & \mbox{\cite{2024ApJ...975..171Y}}& 	8087799 & LC & 19 & Inde. & \mbox{\cite{2017ApJ...850..125Z}} \\
9845907 & SC & 3 & Inde. & \mbox{\cite{2023ApJ...955...80S}}& 	8262223 & SC & - & None & \mbox{\cite{2017ApJ...837..114G}} \\
3440495 & LC & 2 & Sign. & \mbox{\cite{2023ApJ...943L...7L}}& 	11401845 & LC & 15 & Inde. & \mbox{\cite{2017ApJ...835..189L}} \\
3440495 & LC & 25 & Inde. & \mbox{\cite{2022ApJ...937...80M}}& 	6220497 & LC & 15 & Inde. & \mbox{\cite{2016MNRAS.460.4220L}} \\
10417986 & LSC & 1 & Inde. & \mbox{\cite{2022RAA....22j5005F}}& 	4739791 & LC & 6 & Inde. & \mbox{\cite{2016AJ....151...25L}} \\
6382916 & LC & 3 & Inde. & \mbox{\cite{2022ApJ...938L..20N}}& 	9244992 & LC & 28 & Sign. & \mbox{\cite{2015MNRAS.447.3264S}} \\
6382916 & SC & 3 & Inde. & \mbox{\cite{2013MNRAS.433..394U}}& 	8569819 & LC & 1 & Inde. & \mbox{\cite{2015MNRAS.446.1223K}} \\
2857323 & LC & 4 & Inde. & \mbox{\cite{2022ApJ...936...48Y}}& 	7106205 & LC & 2 & Sign. & \mbox{\cite{2015EPJWC.10106013B}} \\
1573174 & LC & 25 & Inde. & \mbox{\cite{2022ApJ...932...42L}}& 	7106205 & LC & 5 & Inde. & \mbox{\cite{2014MNRAS.444.1909B}} \\
5768203 & LC & 2 & Inde. & \mbox{\cite{2022AJ....164...22M}}& 	10080943 & LC & 194 & Inde. & \mbox{\cite{2015A&A...584A..35S}} \\
6951642 & LC & 44 & Inde. & \mbox{\cite{2022A&A...667A..60S}}& 	9533489 & LSC & 14 & Inde. & \mbox{\cite{2015A&A...581A..77B}} \\
5197256 & SC & 20 & Inde. & \mbox{\cite{2021RAA....21..224L}}& 	5892969 & LC & 12 & Sign. & \mbox{\cite{2015A&A...579A.133B}} \\
5197256 & SC & 9 & Sign. & \mbox{\cite{2015JAVSO..43...40T}}& 	11145123 & LC & 9 & Sign. & \mbox{\cite{2014MNRAS.444..102K}} \\
9773821 & LC & 5 & Sign. & \mbox{\cite{2021MNRAS.505.2336M}}& 	8054146 & SC & - & None & \mbox{\cite{2014ApJ...783...89B}} \\
5950759 & LSC & 11 & Inde. & \mbox{\cite{2021MNRAS.504.4039B}}& 	8054146 & SC & 43 & Inde. & \mbox{\cite{2012ApJ...759...62B}} \\
5950759 & LC & 2 & Inde. & \mbox{\cite{2018ApJ...863..195Y}}& 	9764965 & LSC & 22 & Inde. & \mbox{\cite{2014AN....335..812R}} \\
12602250 & LC & 2 & Inde. & \mbox{\cite{2021AJ....162...48L}}& 	9764965 & LSC & 2 & Sign. & \mbox{\cite{2011A&A...535A...3S}} \\
10975348 & SC & 3 & Inde. & \mbox{\cite{2021AJ....161...27Y}}& 	3858884 & SC & 33 & Inde. & \mbox{\cite{2014A&A...563A..59M}} \\
8504570 & SC & 5 & Inde. & \mbox{\cite{2020Galax...8...75L}}& 	4544587 & SC & - & None & \mbox{\cite{2013MNRAS.434..925H}} \\
12268220 & LSC & 15 & Inde. & \mbox{\cite{2020ApJ...898..136C}}& 	11754974 & SC & 27 & Inde. & \mbox{\cite{2013MNRAS.432.2284M}} \\
10736223 & SC & 1 & Inde. & \mbox{\cite{2020ApJ...895..136C}}& 	5988140 & LC & 10 & Inde. & \mbox{\cite{2013A&A...549A.104L}} \\
9850387 & LC & 21 & Inde. & \mbox{\cite{2020ApJ...895..124Z}}& 	4840675 & SC & 22 & Sign. & \mbox{\cite{2012MNRAS.424.1187B}} \\
6629588 & SC & 7 & Inde. & \mbox{\cite{2020AcA....70..265L}}& 	4150611 & LSC & 4 & Sign. & \mbox{\cite{2012MNRAS.422..738S}} \\
10686876 & SC & 4 & Inde. & \mbox{\cite{2020A&A...642A..91L}}& 	10661783 & SC & 25 & Inde. & \mbox{\cite{2011MNRAS.414.2413S}} \\
8975515 & LC & 107 & Inde. & \mbox{\cite{2020A&A...638A..57S}}& 	9700322 & SC & 13 & Inde. & \mbox{\cite{2011MNRAS.414.1721B}} \\
9204718 & LC & 2 & Inde. & \mbox{\cite{2019NewA...71...33U}}& 	11402951 & LSC & 4 & Sign. & \mbox{\cite{2011A&A...535A...3S}} \\
10684673 & SC & 10 & Sign. & \mbox{\cite{2019NewA...67...40T}}& 	11445913 & LSC & 1 & Sign. & \mbox{\cite{2011A&A...535A...3S}} \\
7368103 & LC & 14 & Inde. & \mbox{\cite{2019MNRAS.486.2462W}}& 	8881697 & LSC & 2 & Sign. & \mbox{\cite{2011A&A...535A...3S}} \\

\bottomrule
\end{tabular}
\parbox{\textwidth}{
$\textbf{Note}$. 
The references corresponding to the 'Ref.' column are provided below: \citet{2025PASJ...77..118L},  	\citet{2019MNRAS.486.2129D}, \citet{2024NewA..11302294S},  	\citet{2019ApJ...885...46G}, \citet{2024NewA..11302294S},  	\citet{2018MNRAS.476.4840B}, \citet{2012MNRAS.427.1418M},  	\citet{2019ApJ...884..165Z}, \citet{2024MNRAS.533.2705J},  	\citet{2019ApJ...879...59Y}, \citet{2020A&A...642A..91L},  	\citet{2019AJ....158...88A}, \citet{2024MNRAS.532.1140D},  	\citet{2018MNRAS.475..359B}, \citet{2012MNRAS.419.3028B},  	\citet{2018AcA....68..425S}, \citet{2024MNRAS.527.4052J},  	\citet{2018A&A...616A.130L}, \citet{2016ApJ...826...69G},  	\citet{2017MNRAS.467.4663S}, \citet{2024ApJ...977..241W},  	\citet{2017ApJ...851...39G}, \citet{2024ApJ...977...47S},  	\citet{2017ApJ...850..125Z}, \citet{2024ApJ...975..171Y},  	\citet{2017ApJ...850..125Z}, \citet{2023ApJ...955...80S},  	\citet{2017ApJ...837..114G}, \citet{2023ApJ...943L...7L},  	\citet{2017ApJ...835..189L}, \citet{2022ApJ...937...80M},  	\citet{2016MNRAS.460.4220L}, \citet{2022RAA....22j5005F},  	\citet{2016AJ....151...25L}, \citet{2022ApJ...938L..20N},  	\citet{2015MNRAS.447.3264S}, \citet{2013MNRAS.433..394U},  	\citet{2015MNRAS.446.1223K}, \citet{2022ApJ...936...48Y},  	\citet{2015EPJWC.10106013B}, \citet{2022ApJ...932...42L},  	\citet{2014MNRAS.444.1909B}, \citet{2022AJ....164...22M},  	\citet{2015A&A...584A..35S}, \citet{2022A&A...667A..60S},  	\citet{2015A&A...581A..77B}, \citet{2021RAA....21..224L},  	\citet{2015A&A...579A.133B}, \citet{2015JAVSO..43...40T},  	\citet{2014MNRAS.444..102K}, \citet{2021MNRAS.505.2336M},  	\citet{2014ApJ...783...89B}, \citet{2021MNRAS.504.4039B},  	\citet{2012ApJ...759...62B}, \citet{2018ApJ...863..195Y},  	\citet{2014AN....335..812R}, \citet{2021AJ....162...48L},  	\citet{2011A&A...535A...3S}, \citet{2021AJ....161...27Y},  	\citet{2014A&A...563A..59M}, \citet{2020Galax...8...75L},  	\citet{2013MNRAS.434..925H}, \citet{2020ApJ...898..136C},  	\citet{2013MNRAS.432.2284M}, \citet{2020ApJ...895..136C},  	\citet{2013A&A...549A.104L}, \citet{2020ApJ...895..124Z},  	\citet{2012MNRAS.424.1187B}, \citet{2020AcA....70..265L},  	\citet{2012MNRAS.422..738S}, \citet{2020A&A...642A..91L},  	\citet{2011MNRAS.414.2413S}, \citet{2020A&A...638A..57S},  	\citet{2011MNRAS.414.1721B}, \citet{2019NewA...71...33U},  	\citet{2011A&A...535A...3S}, \citet{2019NewA...67...40T},  	\citet{2011A&A...535A...3S}, \citet{2019MNRAS.486.2462W},  	\citet{2011A&A...535A...3S}.
}
\end{table}

\begin{figure}[p]
\includegraphics[width=\textwidth]{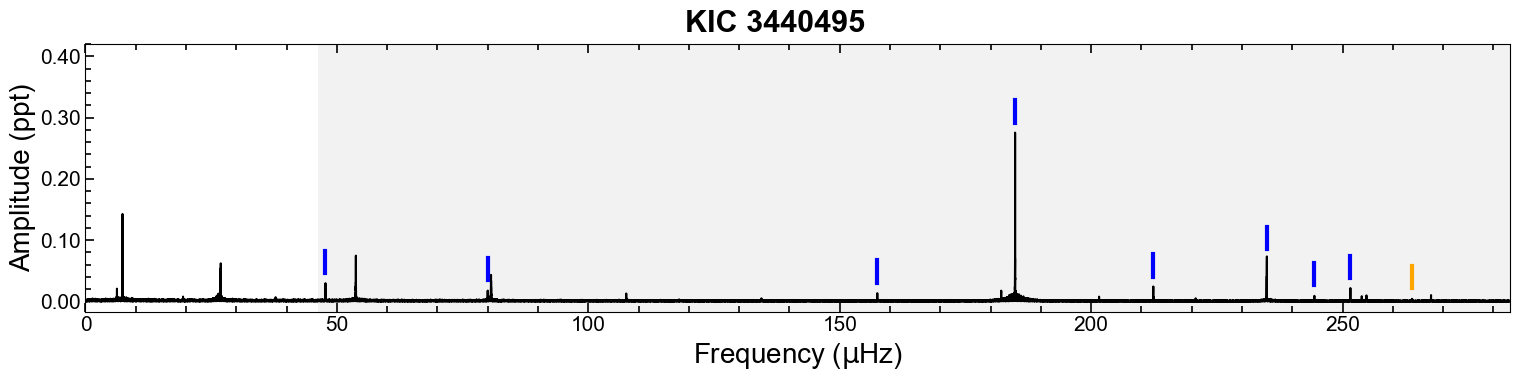}
\includegraphics[width=\textwidth]{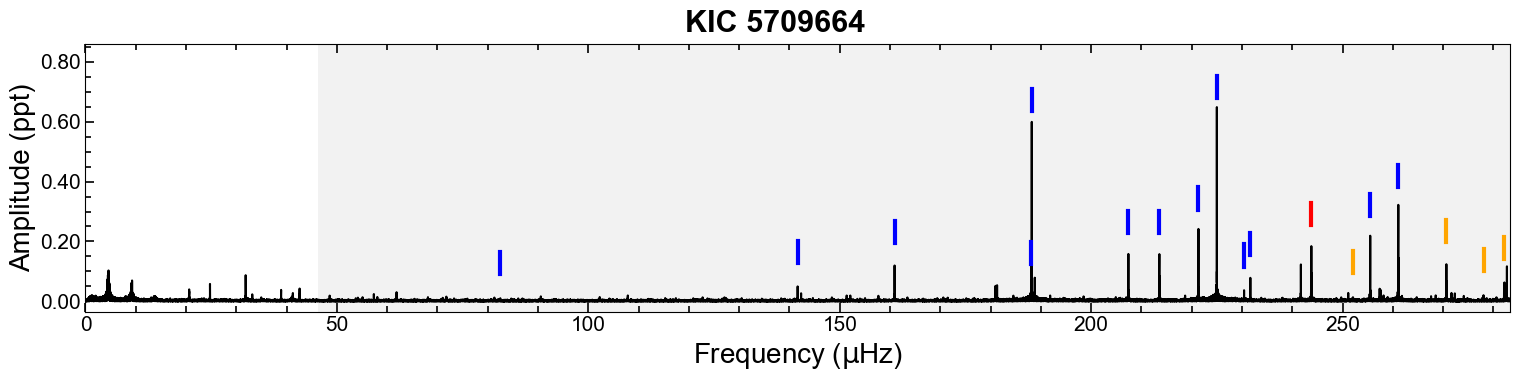}
\includegraphics[width=\textwidth]{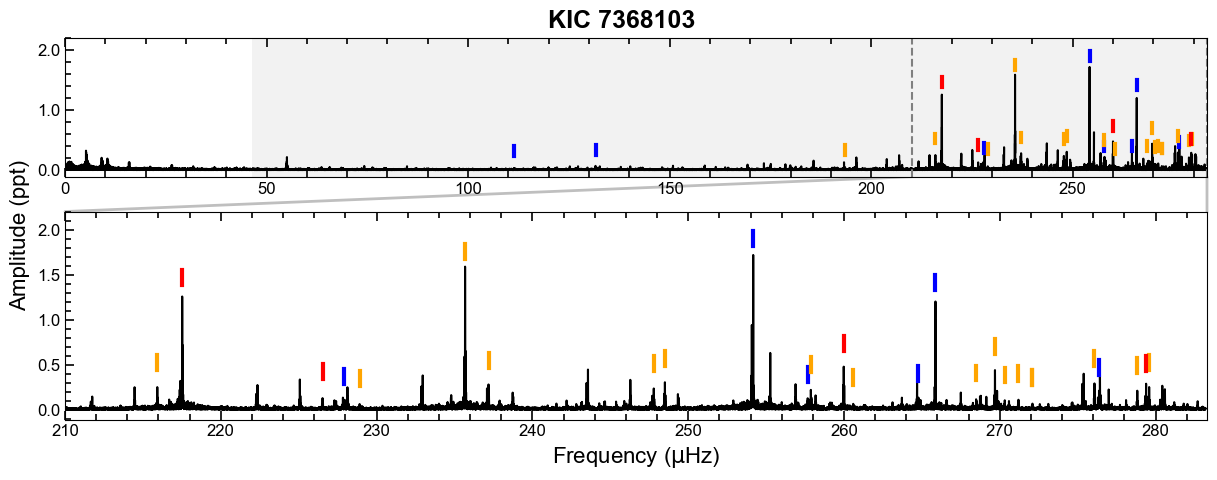}
\includegraphics[width=\textwidth]{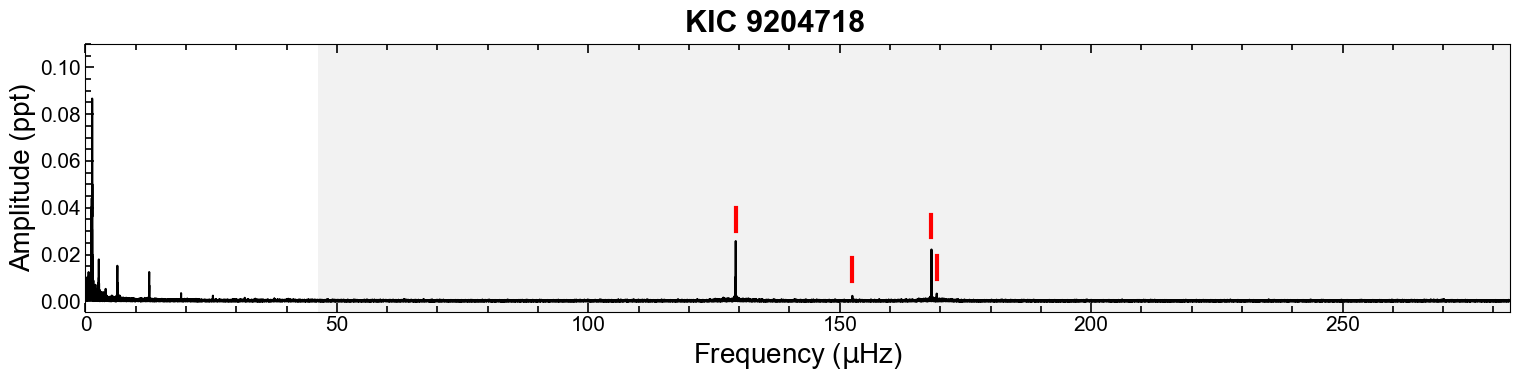}
\caption{Amplitude spectra of the light curves of the four stars. The gray region indicates the p-mode domain above 46 $\upmu$Hz. The genuine independent frequencies (blue), the $F_{\text{rsnf}}$ (red), and the combination frequencies with the SNF modulation (orange) are marked within this region. 
We add an enlarged view for the dense mode distribution of KIC 7368103.
 \label{fig2}}
\end{figure}

\subsection{Detailed Analysis of the Light Curves of the Four Stars}

To ensure the reliability of subsequent frequency extraction, we thoroughly examined the preprocessed light curves of the four target stars.
For the eclipsing binary KIC 7368103, we removed the binary signal.
Using the phase widths of the primary and secondary eclipses from the $Kepler$ Eclipsing Binary Catalog (KEBC \citep{2011AJ....141...83P, 2014AJ....147...45C, 2016AJ....151...68K}), we retained only the out-of-eclipse data for further analysis.

Then, we performed separate frequency analyses for the four stars that show previously unidentified $F_{\text{rsnf}}$.
We employed the Felix method \citep{2010A&A...516L...6C, 2016A&A...585A..22Z} to extract significant frequencies with a signal-to-noise ratio (S/N) greater than 5.6 from the pre-processed light curve data.
S/N is calculated by dividing the frequency amplitude by the median background noise in its surrounding Fourier spectrum \citep{2021ApJ...921...37Z}.
We obtained 76, 253, 326, and 51 frequencies for KIC 3440495, KIC 5709664, KIC 7368103, and KIC 9204718, respectively.
Due to the dense frequency spectra of the $\delta$ Sct stars shown in Figure \ref{fig2}, we performed a detailed examination and analysis of the extracted frequencies: 
(1) Frequencies that are separated by less than the frequency resolution $1.5/T$, where $T$ is the time span of the data, are treated as a single frequency to avoid frequency aliasing and ensure the accuracy of the analysis. 
(2) We then used the sLSP method (window = 300d, step = 5d) to check all frequencies and find all $F_{\text{rsnf}}$ candidates. 
There are a total of 32 $F_{\text{rsnf}}$ candidates, and all modulation patterns are shown in Figure \ref{fig3}. 
(3) Using the method outlined in \citet{2012AN....333.1053P}, we  examined and identified harmonic frequencies $f = nf_{i} \pm {1.5/T}$ and combination frequencies $f = mf_{i} + nf_{j} + lf_{k} \pm {1.5/T}$. 
These frequencies are typically artifacts arising from nonlinear mode interactions and should be removed to obtain a clean set of intrinsic pulsation frequencies.
The pulsation frequencies of the four $\delta$ Sct stars are listed in Table \ref{tab2}, which includes only independent frequencies and all $F_{\text{rsnf}}$ candidates.
The frequency IDs are sorted in descending order of amplitude as $f_{\text{1}}$, $f_{\text{2}}$, etc.

$F_{\text{rsnf}}$ may result from either a single or multiple reflections of a SNF.
Although there is no theoretical limit to the number of reflections, practical detectability is constrained by noise and frequency resolution \citep{2013MNRAS.430.2986M}.
Considering the pulsation frequency range of $\delta$ Sct stars, as shown in Fig.2 of \citet{2025A&A...693A..63W}, $F_{\text{rsnf}}$ can be obtained by reflecting a SNF (within the range of $1 \times f_{ny}$ to $2 \times f_{ny}$) once, or by reflecting a SNF (greater than $2 \times f_{ny}$) twice.
Therefore, we checked whether the two SNFs corresponding to each independent $F_{\text{rsnf}}$ are combination frequencies.
In the `Label' column of Table \ref{tab2}, we categorize the types of frequencies: 
(1) Independent frequencies that are not candidates for $F_{\text{rsnf}}$ are labeled as `Base'.
(2) If both the $F_{\text{rsnf}}$ candidate and its corresponding SNFs are independent frequencies, they are considered as real $F_{\text{rsnf}}$ and labeled as `$F_{\text{rsnf}}$'.
(3) If the $F_{\text{rsnf}}$ candidate is a combination frequency, it is not considered as a real $F_{\text{rsnf}}$ and is labeled as `Combination'.
(4) If the $F_{\text{rsnf}}$ candidate is an independent frequency but at least one of its corresponding SNFs is a combination frequency, it is not considered as a real $F_{\text{rsnf}}$ and is labeled as ` Mirror', with the corresponding SNF value noted in the `Remark' column.
In Figure \ref{fig2}, frequencies above 46 $\upmu$Hz are highlighted with a gray background and color-coded by category. 
An enlarged view is provided for KIC 7368103 due to its relatively dense frequency distribution.




\begin{table}[ht!]
\caption{Pulsation frequencies of the four stars extracted from the $Kepler$ LC light curves. The 'S/N' column represents the signal-to-noise ratio.}\label{tab2}
\begin{tabular}{
                >{\centering\arraybackslash}p{0.5cm} 
                >{\centering\arraybackslash}p{2.6cm} 
                >{\centering\arraybackslash}p{2.5cm} 
                >{\centering\arraybackslash}p{1.5cm}
                >{\centering\arraybackslash}p{2cm}
                >{\centering\arraybackslash}p{2.2cm}} 

\toprule
\multicolumn{6}{c}{KIC 3440495 \citep{2022ApJ...937...80M}} \\
\hline
ID & Frequency($\upmu$Hz) & Amplitude(ppt) & S/N & Label & Remark \\
\hline

$f_{\text{1}}$ & 26.8930(2) & 0.0337(1) & 231.77 & $f_{\text{rot}}$ & $f_{\text{rot}}$* \\
$f_{\text{2}}$ & 184.876106(6) & 0.02752(4) & 704.39 & Base & $F_{\text{0}}$* \\
$f_{\text{3}}$ & 7.36662(3) & 0.0144(1) & 137.64 & Base & * \\
$f_{\text{4}}$ & 234.94726(2) & 0.00727(3) & 213.97 & Base & $F_{\text{1}}$* \\
$f_{\text{5}}$ & 47.7275(1) & 0.00300(10) & 31.18 & Base & * \\
$f_{\text{6}}$ & 212.36676(6) & 0.00243(3) & 70.17 & Base & * \\
$f_{\text{7}}$ & 251.55782(7) & 0.00213(3) & 62.71 & Base & * \\
$f_{\text{8}}$ & 79.9938(2) & 0.00188(9) & 20.58 & Base & * \\
$f_{\text{9}}$ & 6.2618(3) & 0.0018(1) & 16.78 & Base & * \\
$f_{\text{10}}$ & 157.4735(2) & 0.00133(5) & 28.66 & Base & * \\
$f_{\text{11}}$ & 26.5801(6) & 0.0010(1) & 6.89 & Base & * \\
$f_{\text{12}}$ & 244.4004(2) & 0.00086(4) & 24.52 & Base & * \\
- & 263.8190(3) & 0.00040(3) & 12.6 & Combination & * \\

\toprule
\multicolumn{6}{c}{KIC 5709664 \citep{2019MNRAS.486.2129D}} \\
\hline
ID & Frequency($\upmu$Hz) & Amplitude(ppt) & S/N & Label & Remark \\
\hline

$f_{\text{1}}$ & 4.58575(2) & 0.0985(5) & 199.87 & Base & \\
$f_{\text{2}}$ & 225.000644(5) & 0.06894(8) & 816.31 & Base & * \\
$f_{\text{3}}$ & 188.189826(5) & 0.06412(7) & 905.74 & Base & * \\
$f_{\text{4}}$ & 261.088237(10) & 0.03455(8) & 443.75 & Base & * \\
$f_{\text{5}}$ & 221.34095(1) & 0.02604(8) & 307.73 & Base & * \\
$f_{\text{6}}$ & 255.50239(2) & 0.02391(8) & 288.25 & Base & * \\
$f_{\text{7}}$ & 243.79153(2) & 0.02020(8) & 253.44 & $F_{\text{rsnf}}$ & ** \\
$f_{\text{8}}$ & 207.406280(2) & 0.01764(10) & 178.13 & Base & * \\
$f_{\text{9}}$ & 213.56893(2) & 0.01690(8) & 223.12 & Base & * \\
$f_{\text{10}}$ & 82.49272(2) & 0.01519(7) & 227.8 & Base & \\
- & 270.64660(3) & 0.01372(8) & 165.48 & Mirror & 837.08178** \\
$f_{\text{11}}$ & 160.91010(2) & 0.01300(6) & 204.87 & Base & * \\
$f_{\text{12}}$ & 4.4185(2) & 0.0093(5) & 18.44 & Base & \\
$f_{\text{13}}$ & 231.66074(4) & 0.00893(8) & 108.7 & Base & * \\
- & 282.18947(9) & 0.0067(1) & 45.94 & Combination & ** \\
$f_{\text{14}}$ & 141.63150(5) & 0.00559(6) & 90.79 & Base & * \\
$f_{\text{15}}$ & 20.6230(1) & 0.0046(1) & 32.53 & Base & \\
$f_{\text{16}}$ & 230.41461(9) & 0.00423(8) & 50.35 & Base & * \\
$f_{\text{17}}$ & 188.06807(4) & 0.00378(7) & 53.42 & Base & \\
- & 278.0299(2) & 0.00216(10) & 21.64 & Combination & ** \\
- & 252.0693(2) & 0.00179(8) & 21.83 & Combination & ** \\

\hline
\end{tabular}
\parbox{\textwidth}{
$\textbf{Note}$. 
'$F_{\text{rsnf}}$' denotes an independent $F_{\text{rsnf}}$.
'Base' indicates a real independent frequency.
'Combination' marks a $F_{\text{rsnf}}$ that is a combination frequency.
'Mirror' marks an independent $F_{\text{rsnf}}$ that corresponds to a combination SNF. 
* indicates that the frequency was reported in the literature as a true frequency within the Nyquist range. 
** indicates that the corresponding SNF component was reported in the literature.
*** indicates that the frequency was reported on both sides of $f_{\text{ny}}$.
${f_{\text{orb}}}^*$ , ${f_{\text{rot}}}^*$, ${F_0}^*$, and ${F_1}^*$ represent the orbital frequency, the rotational frequency, the fundamental frequency, and the first-overtone, respectively, as identified in the literature.
}
\end{table}

\begin{table}[ht!]
\caption*{\textbf{Table 2.} Continued.}
\begin{tabular}{
                >{\centering\arraybackslash}p{0.5cm} 
                >{\centering\arraybackslash}p{2.6cm} 
                >{\centering\arraybackslash}p{2.5cm} 
                >{\centering\arraybackslash}p{1.5cm}
                >{\centering\arraybackslash}p{2cm}
                >{\centering\arraybackslash}p{2.2cm}}
\toprule
\multicolumn{6}{c}{KIC 7368103 \citep{2019MNRAS.486.2462W}} \\
\hline
ID & Frequency($\upmu$Hz) & Amplitude(ppt) & S/N & Label &Remark \\
\hline

$f_{\text{1}}$ & 254.180732(2) & 2.1009(8) & 2661.43 & Base & *** \\
$f_{\text{2}}$ & 265.880055(6) & 0.5573(7) & 747.39 & Base & *** \\
$f_{\text{3}}$ & 264.781708(9) & 0.3927(8) & 508.48 & Base & * \\
$f_{\text{4}}$ & 276.374043(9) & 0.3468(7) & 494.87 & Base & ** \\
$f_{\text{5}}$ & 217.52977(1) & 0.3419(9) & 395.69 & $F_{\text{rsnf}}$ & ** \\
$f_{\text{6}}$ & 259.98341(1) & 0.2957(8) & 363.71 & $F_{\text{rsnf}}$ & ** \\
$f_{\text{7}}$ & 257.68640(2) & 0.2229(9) & 250.19 & Base & ** \\
$f_{\text{8}}$ & 227.90604(1) & 0.2103(7) & 306.01 & Base & \\
- & 235.67922(2) & 0.1462(8) & 186.56 & Mirror & 330.7560** \\
$f_{\text{9}}$ & 226.53240(2) & 0.1260(7) & 185.28 & $F_{\text{rsnf}}$ & \\
$f_{\text{10}}$ & 279.39279(4) & 0.0755(7) & 111.05 & $F_{\text{rsnf}}$ & ** \\
- & 269.69148(7) & 0.0445(7) & 62.45 & Combination & *** \\
- & 248.50633(8) & 0.0351(7) & 53.52 & Combination & ** \\
$f_{\text{13}}$ & 5.2144(4) & 0.031(3) & 10.03 & Base & \\
- & 276.0763(1) & 0.0297(7) & 42.01 & Mirror & 290.3589** \\
- & 237.1843(1) & 0.0288(8) & 37.62 & Mirror & 329.2509** \\
- & 279.5865(1) & 0.0252(7) & 37.39 & Combination & ** \\
- & 257.8725(2) & 0.0241(9) & 26.86 & Mirror & 308.5627** \\
- & 215.9230(2) & 0.0208(8) & 25.03 & Combination & ** \\
$f_{\text{14}}$ & 9.0877(5) & 0.020(2) & 8.69 & Base & \\
- & 278.8247(2) & 0.0182(7) & 24.64 & Combination & ** \\
- & 247.7908(2) & 0.0157(7) & 23.57 & Combination & ** \\
- & 271.1985(2) & 0.0118(6) & 18.83 & Combination & ** \\
$f_{\text{15}}$ & 111.3642(1) & 0.0112(4) & 30.48 & Base & \\
- & 270.3188(3) & 0.0110(7) & 15.41 & Combination & ** \\
- & 260.5924(4) & 0.0091(8) & 11.68 & Combination & \\
- & 268.4792(4) & 0.0077(8) & 9.9 & Combination & ** \\
- & 228.9096(4) & 0.0068(6) & 10.56 & Combination & \\
- & 272.0965(5) & 0.0050(6) & 8.33 & Combination & \\
- & 193.5270(4) & 0.0046(5) & 10.04 & Combination & \\
- & 249.3348(7) & 0.0039(7) & 5.93 & Combination & \\
$f_{\text{16}}$ & 131.6350(4) & 0.0043(4) & 10.8 & Base & \\

\toprule
\multicolumn{6}{c}{KIC 9204718 \citep{2019NewA...71...33U}} \\
\hline
ID & Frequency($\upmu$Hz) & Amplitude(ppt) & S/N & Label & Remark \\
\hline

$f_{\text{1}}$ & 2.66288(2) & 0.0257(1) & 220.74 & Base & \\
$f_{\text{2}}$ & 1.32810(7) & 0.0103(2) & 61.09 & Base & * \\
$f_{\text{3}}$ & 1.1698(2) & 0.0047(2) & 26.66 & Base & \\
$f_{\text{4}}$ & 129.31655(3) & 0.00270(2) & 131.36 & $F_{\text{rsnf}}$ & * \\
$f_{\text{5}}$ & 168.24455(4) & 0.00216(2) & 105.16 & $F_{\text{rsnf}}$ & * \\
$f_{\text{6}}$ & 6.3357(1) & 0.00162(6) & 29.18 & $f_{\text{orb}}$ & $f_{\text{orb}}$* \\
$f_{\text{7}}$ & 169.2904(2) & 0.00036(2) & 17.66 & $F_{\text{rsnf}}$ & \\
$f_{\text{8}}$ & 152.5085(3) & 0.00024(2) & 14.19 & $F_{\text{rsnf}}$ & \\

\hline
\end{tabular}
\end{table}

\section{Discussion}

\begin{figure}[ht]
\includegraphics[width=\textwidth]{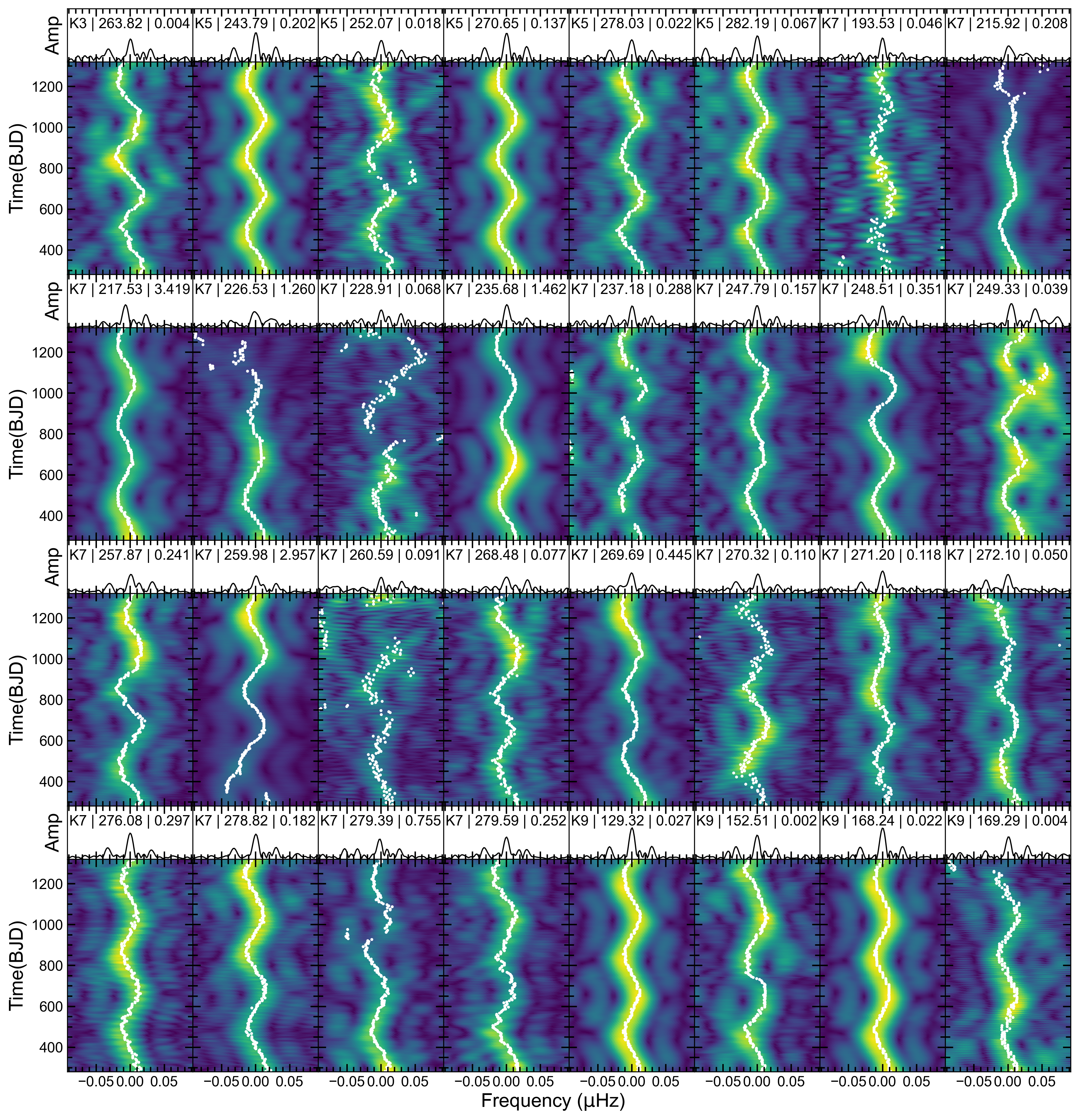}
\caption{A total of 32 SNF modulations of the four stars are shown here.
The figure sequentially labels the KIC ID, frequency ($\upmu$Hz), and amplitude (ppt) for each panel, separated by '|'.
'K3', 'K5', 'K7', and 'K9' represent KIC 3440495, KIC 5709664, KIC 7368103, and KIC 9204718, respectively. 
Color scale is the same as in Figure \ref{fig1}.
\label{fig3}}
\end{figure}

Since the frequency datasets from previous studies were processed by the original authors, it is not possible to determine whether any $F_{\text{rsnf}}$ were present in the raw data, or to quantify their number.
However, we extracted a total of 706 frequencies from four stars, among which 32 were identified as $F_{\text{rsnf}}$ candidates using the sLSP method, corresponding to 4.53$\%$—significantly higher than the 0.67$\%$ reported by \citet{2025A&A...693A..63W} for $\gamma$ Doradus stars.
The modulation patterns of these $F_{\text{rsnf}}$ candidates are shown in Figure \ref{fig3}, and none of them fall within the typical g-mode pulsation range.
17 of these candidates fall within the range of 250 $\upmu$Hz to the $f_{\text{ny}}$ (283.16 $\upmu$Hz), accounting for 8.7$\%$, slightly higher than the 7$\%$ reported by \citet{2025A&A...693A..63W}.
These findings further support the conclusion that aliasing between $F_{\text{rsnf}}$ and real frequencies is more likely to occur within the p-mode pulsation domain of $\delta$ Sct stars.

As shown in the 'Remark' column of Table \ref{tab2}, we identified additional frequencies that were not reported in the corresponding studies.
This discrepancy may result from subtle differences in data preprocessing, variations in the adopted S/N thresholds, or minor deviations affecting the identification of combination frequencies. 
These differences are typical in frequency extraction and combination analysis and do not affect the reliability of the main conclusions.

In the study of KIC 3440495, \citet{2022ApJ...937...80M} removed $F_{\text{rsnf}}$ candidate using the $Kepler$ SC photometric data. 
However, they retained the frequency of 263.82 $\upmu$Hz. 
This frequency shows SNF modulation but corresponds to a combination frequency in our analyses.
In Figure \ref{fig1}, the panel for KIC 5709664 shows 282.11 $\upmu$Hz as the central frequency, while the SNF modulation we identified appears at a different location within the same figure. 
The slight offset suggests they may represent different frequencies, with our $F_{\text{rsnf}}$ candidate lying near 282.19 $\upmu$Hz, which we classify as a combination frequency. 
It is likely that this candidate was recognized and removed by \citet{2019MNRAS.486.2129D} during their analysis.
282.11 $\upmu$Hz is also extracted in our analysis but corresponds to a combination frequency.
Among the frequencies reported for KIC 7368103 by \citet{2019MNRAS.486.2462W}, two frequencies—217.50 $\upmu$Hz and 269.68 $\upmu$Hz—show signs of SNF modulation. 
Our analysis identified 217.50 $\upmu$Hz as a real $F_{\text{rsnf}}$, while 269.68 $\upmu$Hz corresponds to a combination frequency.
In the study of KIC 9204718 by \citet{2019NewA...71...33U}, the reported frequencies 129.32 $\upmu$Hz and 168.24 $\upmu$Hz are classified in this work as real $F_{\text{rsnf}}$.
Additionally, we identified 1 real $F_{\text{rsnf}}$ in the data from KIC 5709664, 3 in KIC 7368103 and 2 in KIC 9204718, which may have been discarded in previous studies.
A complete list of these frequencies is given in Table \ref{tab3}.

For asteroseismic modeling, only the filtered and validated independent frequencies (see Table \ref{tab2}) would be used.
Among the four stars, KIC 3440495 had 12 independent frequencies that were all confirmed as genuine. 
For the other three stars, the fractions of $F_{\text{rsnf}}$ among the independent frequencies were 1/17 for KIC 5709664, 4/16 for KIC 7368103, and 4/8 for KIC 9204718.

The aliasing of $F_{\text{rsnf}}$ with the real frequencies has long been recognized as a potential source of misleading results in frequency analysis and frequency-based studies.
The charts clearly demonstrate that $F_{\text{rsnf}}$ exhibit multi-peak structures, which makes the spectral profile filtering an effective, yet incomplete, method for their removal.
In cases of poor data quality, $F_{\text{rsnf}}$ may not be fully detectable through profile identification.
The new sLSP method, however, offers the capability for batch searching, enabling the identification of $F_{\text{rsnf}}$ with less obvious multi-peak structures.

\begin{table}[ht] 
\caption{Real $F_{\text{rsnf}}$ and reference $F_{\text{rsnf}}$. \label{tab3}}
\begin{tabular}{>{\centering\arraybackslash}p{2.8cm} 
                >{\centering\arraybackslash}p{2.7cm} 
                >{\centering\arraybackslash}p{2.6cm} 
                >{\centering\arraybackslash}p{3.6cm}}
\toprule
KIC ID & Real $F_{\text{rsnf}}$ ($\upmu$Hz) & Reference $F_{\text{rsnf}}$ ($\upmu$Hz) & Type\\
\midrule
KIC 3440495 & - & 263.82  & Combination\\
\midrule
KIC 5709664 & 243.79153(2) & - & $F_{\text{rsnf}}$ \\
KIC 5709664 & 282.18947(9) & 282.11 & Combination \\
\midrule
KIC 7368103 & 217.52977(1) & 217.50  & $F_{\text{rsnf}}$ \\
KIC 7368103 & 226.53240(2) & - & $F_{\text{rsnf}}$ \\
KIC 7368103 & 259.98341(1) & - & $F_{\text{rsnf}}$ \\
KIC 7368103 & 269.69148(7) & 269.68  & Combination \\
KIC 7368103 & 279.39279(4) & - & $F_{\text{rsnf}}$ \\
\midrule
KIC 9204718 & 129.31655(3) & 129.32  & $F_{\text{rsnf}}$ \\
KIC 9204718 & 168.24455(4)  & 168.24  & $F_{\text{rsnf}}$ \\
KIC 9204718 & 169.2904(2)  & - & $F_{\text{rsnf}}$ \\
KIC 9204718 & 152.5085(3)  & - & $F_{\text{rsnf}}$ \\
\bottomrule
\end{tabular}

\parbox{\textwidth}{
$\textbf{Note}$. 
'Reference $F_{\text{rsnf}}$' is the $F_{\text{rsnf}}$ we first found in the literature. 
}
\end{table}

\section{Summary}

We conducted a review of 74 published studies during the past ten years that investigated the pulsation behavior of $\delta$ Sct stars using the $Kepler$ data, encompassing 68 individual stars. 
For each publication, we recorded the number of reported significant frequencies falling within the range of 46 $\upmu$Hz to $f_{\text{ny}}$ of the LC data, corresponding to the p-mode frequency domain where $F_{\text{rsnf}}$ is most likely to undergo aliasing.
We then downloaded the $Kepler$ light curves, performed preprocessing, and re-examined the 1,406 reported frequencies using the sLSP method \citep{2025A&A...693A..63W}.

As a result, we identified four stars—KIC 3440495, KIC 5709664, KIC 7368103, and KIC 9204718—whose published frequencies exhibit clear signs of SNF modulation (Figure \ref{fig1}), which had not been recognized in the original studies.
We subsequently extracted all significant frequencies from the light curves of these four stars and applied the sLSP method, yielding 32 $F_{\text{rsnf}}$ candidates, each of which was carefully examined and classified. 
Table \ref{tab2} summarizes the independent frequencies and $F_{\text{rsnf}}$ candidates for the four stars. 
The corresponding amplitude spectra are presented in Figure \ref{fig2}.

Accurately identifying pulsation frequencies is a fundamental and crucial step in asteroseismic studies. 
Matching a moderate number of well-validated, independent frequencies with asteroseismic model frequencies can better constrain the models, leading to more precise stellar parameters \citep{2021ApJ...920...76C, 2022ApJ...936...48Y}.
Using incorrect frequencies, however, may lead to substantial errors in the resulting models.
$f_{\text{ny}}$ of the $Kepler$ LC data is 283.16 $\upmu$Hz. 
Real p-mode frequencies above this limit are aliased below $f_{\text{ny}}$, overlapping with low-frequency signals and complicating mode identification. 
In principle, SNF artifacts exhibit uniformly spaced multiplet structures. 
However, when their frequency spacing mimics that of rotational splitting, misclassification may occur \citep{2013MNRAS.430.2986M}.
Removing harmonic and combination frequencies, combining SC and LC data, and integrating techniques such as mode identification and frequency spacing analysis constitute the most effective approach currently available for more accurately identifying real frequencies.

Our results further demonstrate that using SC data or identifying multiple structures in the power spectra to mitigate SNF contamination still has limitations. 
The sLSP method proves to be a robust tool for verifying the presence of $F_{\text{rsnf}}$. 
When treated as an independent phenomenon, SNF modulation serves as a valuable tool for improving the accuracy of asteroseismic analysis.
Identifying a sufficiently complete set of $F_{\text{rsnf}}$ not only facilitates removal spurious frequencies, but also enables the recovery of the corresponding true SNFs. 
This process leads to a more reliable and complete frequency list , thereby improving the constraints on asteroseismic models. 
Its significance is more evident when $F_{\text{rsnf}}$ account for a substantial portion of the independent frequency set—for example, in KIC 9204718, half of the independent frequencies are identified as $F_{\text{rsnf}}$. 
The influence of $F_{\text{rsnf}}$ on asteroseismic modeling will be further evaluated in our subsequent modeling efforts.

In future work, we aim to apply the sLSP method to a broader range of $\delta$ Sct stars and extend our analysis to other types of pulsating variables, such as B-type pulsators \citep{2023ApJ...953....9Z, 2023A&A...679A...6M}.
This expanded effort will contribute to advancing nonlinear stellar oscillation theory and inform stellar modeling approaches that can interpret the pulsation behavior of variable stars. 
Meanwhile, we look forward to higher-quality data from missions such as PLATO \citep{2025ExA....59...26R}, offering 2.5-second exposure observations, to better study pulsating stars.

\vspace{6pt} 



\authorcontributions{
 Conceptualization, W.Z.; data curation, X.W. and Y.M.; methodology, X.W. and Y.M.; writing---original draft preparation, Z.Y.; writing---review and editing, J.F. and W.Z.
All authors have read and agreed to the published version of the manuscript.
}

\funding{
This work is supported by the China Manned Space Program with grant no. CMS-CSST-2025-A013, and the Central Guidance for Local Science and Technology Development Fund under No. ZYYD2025QY27.

}

\dataavailability{
The data underlying this article will be shared on reasonable request
to the corresponding author.
}

\acknowledgments{
We acknowledge the support from the National Natural Science Foundation of China (NSFC) through the grants 12090040, 12090042 and 12427804.

}

\conflictsofinterest{The authors declare no conflicts of interest.} 

\begin{adjustwidth}{-\extralength}{0cm}

\reftitle{References}


\bibliographystyle{mdpi}
\bibliography{reference}

\PublishersNote{}
\end{adjustwidth}
\end{document}